\documentclass[a4paper,12pt]{article}

\usepackage{ifpdf}
\newif\ifpdf
\ifx\pdfoutput\undefined
  \pdffalse
\else
  \pdfoutput=1
  \pdftrue
\fi
\RequirePackage{xspace}
\RequirePackage{subfigure} %
\RequirePackage[centertags]{amsmath} %
\RequirePackage{amssymb}
\RequirePackage{wrapfig} %
\RequirePackage{calc} %
\RequirePackage{ifthen}
\RequirePackage{tabularx} %
\RequirePackage{flafter} %
\RequirePackage{fancyhdr} %
\ifpdf
  \RequirePackage[pdftex]{color}%
  \RequirePackage{colortbl}%
  \RequirePackage{array}%
  \RequirePackage[pdftex]{graphicx}
  \RequirePackage[ pdftex, plainpages = false, pdfpagelabels,
                 pdfpagelayout = useoutlines,
                 bookmarks,
                 breaklinks = true,
                 linktocpage,
                 pagebackref,
                 colorlinks = true,
                 linkcolor = blue,
                 urlcolor  = blue,
                 citecolor = blue,
                 anchorcolor = blue,
                 hyperindex = true,
                 hyperfigures
                 ]{hyperref}
\else
  \RequirePackage{color}
  \RequirePackage{colortbl}
   \RequirePackage{array}
  \RequirePackage[dvips]{graphicx}
  \RequirePackage{hyperref}
  \usepackage{rotating}
\fi
\usepackage{makeidx}
\usepackage{setspace}
\usepackage{rotating}
\usepackage{ecltree}
\usepackage{epic}
\usepackage{supertabular}
\usepackage{color}
\usepackage{exscale}
\usepackage{fontenc}
\usepackage{ifthen}
\usepackage{latexsym}
\usepackage{makeidx}
\usepackage{syntonly}
\usepackage{inputenc}
\usepackage{graphicx}
\usepackage{setspace}
\usepackage{caption2}
\usepackage[english]{babel}
\usepackage[square, comma,numbers,sort&compress]{natbib}
\usepackage{hypernat}
\usepackage{boxedminipage}
\usepackage{framed}
\usepackage{longtable}
\usepackage[all]{hypcap}
\newcommand{\note}[1]{\marginpar[left]{\singlespace \tiny #1}}

\renewcommand{\note}[1]{}

\doublespace

\title
{ %
\vspace*{3.0cm} \LARGE{\bf Testing the Connectivity of Networks} \vspace*{4.0cm} \\
}

\author{Taha Sochi\footnote{Imaging Sciences \& Biomedical Engineering, King's College London, The Rayne
Institute, St Thomas' Hospital, London, SE1 7EH, UK. Email: taha.sochi@kcl.ac.uk.} \vspace*{5.0cm}}

\begin{document}

\maketitle %
\pagenumbering{arabic}

\newpage
\phantomsection \addcontentsline{toc}{section}{Contents} %
\tableofcontents

\newpage
\phantomsection \addcontentsline{toc}{section}{List of Figures} %
\listoffigures


\newpage
\phantomsection \addcontentsline{toc}{section}{Abstract} \noindent
{\noindent \LARGE \bf Abstract} \vspace{0.5cm}\\
\noindent %

In this article we discuss general strategies and computer algorithms to test the connectivity of
unstructured networks which consist of a number of segments connected through randomly distributed
nodes.

\vspace{1cm}


\pagestyle{headings} %
\addtolength{\headheight}{+1.6pt}
\lhead[{Chapter \thechapter \thepage}]%
      {{\bfseries\rightmark}}
\rhead[{\bfseries\leftmark}]%
     {{\bfseries\thepage}}
\headsep = 1.0cm

\newpage
\section{Introduction}

\begin{figure}[!h]
\centering{}
\includegraphics
[scale=0.7] {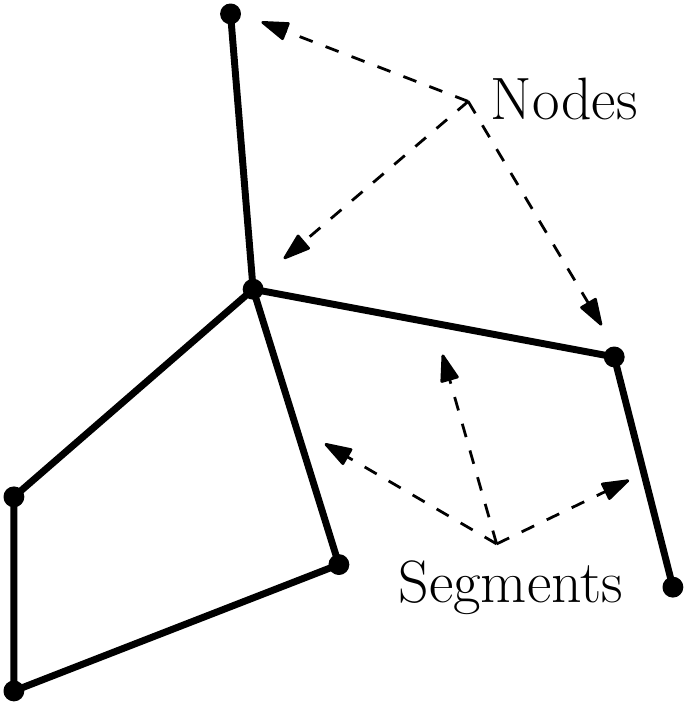} \caption{A simple network.} \label{NetTerm}
\end{figure}

The network, as an abstract concept, consists of segments joined at randomly distributed nodes, as
presented schematically in Figure \ref{NetTerm}. The segments can represent routes, ducts, pipes,
paths, streets, wires, elements in the finite difference or finite element methods, etc., while the
nodes can represent pores, junctions, crossroads, pylons, airports, terminals, and so on. In many
scientific and engineering applications, networks are used as model input data to describe the
topology, and might even the geometry, of certain physical or theoretical objects which are under
investigation. Examples of network include real and model porous media \cite{SorbieCJ1989,
OrenBA1997, TsakiroglouTKPS2003, PerrinTSC2006, Sochithesis2007, SochiB2008}, vascular networks of
blood vessels \cite{SherwinFPP2003, RuanCZC2003, FormaggiaLTV2006, AlastrueyMPDPS2007}, electronic
circuits \cite{RouxH1987, Schilders2009, ZhanMT2009}, fluid transport pipe systems
\cite{RatnayakeJ1999, HammaDF2008, DobersekG2009}, city traffic routes \cite{BentleyL1980,
MessmerP1994, Ledoux1997}, computing and communication networks \cite{SarvothamRB2005,
JiangXCHX2011}, the World Wide Web \cite{SohY1998, BarabasiAJ2000, Sgroi2008}, electric power grids
\cite{CrucittiLM2004, ChassinP2005, LinY2011}, railway nets \cite{HuismanBD2002, MeesterM2007},
decay routes of excited quantum systems such as atomic and molecular species \cite{SochiEmis2010},
and so on.

One condition that is commonly required in these networks is total connectivity, that is the
network must be a single joined entity with no cut or separation. Strictly speaking, the total
connectivity of a network means that any junction in the network can be reached from any other
junction following the routes inside the network without jump. The network is partially connected
if it consists of a number of totally connected partitions.

As most of these networks, especially the large ones, are produced through automated computational
processes, the connectivity of the network cannot be guaranteed. In fact even the manually built
networks can suffer from disconnectivities due to human errors. Because direct tests by human
inspection (e.g. by tracking the routes and building connectivity tables) is not practical except
for the very small networks, automated computer algorithms based on certain strategies are required
to test the connectivity of these networks.

In this article, unless stated otherwise, we deal with general networks without specific condition
on their structure. In this context, some terminology may provide helpful clarification. The
connectivity index, $c$, of a node is the number of segments connected to that node. The
connectivity index, in general, can take the values $c=0$ for a singular non-connected node, $c=1$
for a boundary node which is connected to a single segment, $c=2$ for a bridge node connecting only
two segments, and $c=n>2$ for bifurcation (branching) nodes. The network is unstructured if it has
variable node connectivity index with random node distribution and indexing. The connectivity index
may also be attributed to the network as a whole, in which case it means the average node
connectivity index and is usually computed as twice the total number of segments divided by the
total number of connected nodes which excludes the singular ones.

\begin{figure}[!h]
\centering{}
\includegraphics
[scale=0.7] {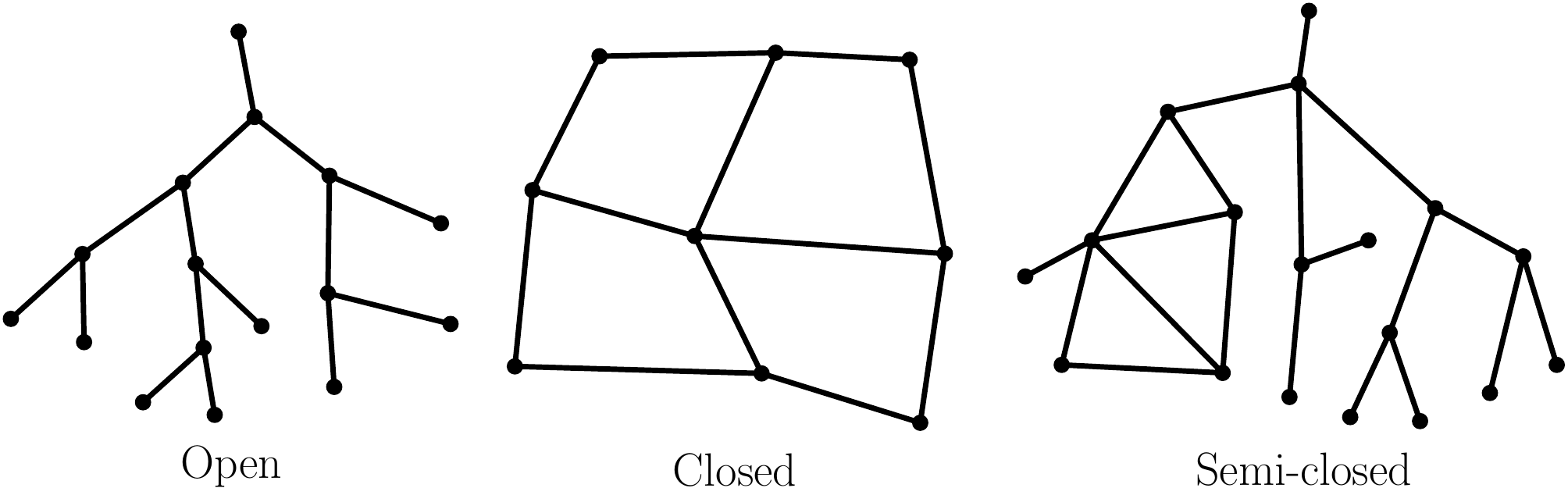} \caption{Schematic representation of network types.} \label{NetType}
\end{figure}

The totally-connected network (as well as each partition of a partially connected network) can be
open where each node can be reached from any other node through a unique non-retraced path, or
closed where each node can be reached from any other node through a number of distinct paths, or
semi-closed where some nodes can be reached through unique paths while others can be reached
through multiple paths. A schematic representation of these three types of network are presented in
Figure \ref{NetType}. The assumption here is that each segment in the network has exactly two nodes
which define its ends. For the open networks, these two nodes uniquely identify the segment and
distinguish it from all other segments. In the following we assume that there is no singular nodes
in the network as they serve no purpose in the context of connectivity and hence can be removed
from the network.

\section{Methods}

There are many methods for testing the connectivity of networks. In the following we outline some
of these methods with proposed algorithms and graphic illustrations. As we do not assume a specific
topology for the networks, the algorithms and conclusions are general.

\subsection{Direct Inspection}

Direct connectivity inspection method is the most straight test to verify network connectivity
since it is derived from the definition of total connectivity, that is in a totally connected
network each node is connected to all other nodes in the network and hence any node can be reached
from any other node within the network routes. The requirement, therefore, of this simple method is
to find out through a systematic route inspection if all other nodes can be reached from each node
in the network. This can be achieved, for example, by constructing a two-dimensional square array
where both dimensions represent the list of all nodes in the network (e.g rows for origin nodes and
columns for destination nodes) and the entries in the array can represent Boolean flags for finding
a connecting route between the corresponding nodes representing the row and column of the
particular entry. The network is then totally connected if all the entries, after a systematic
route inspection, are found to be true.

However, because connectivity is a reflexive relation (i.e. each node is connected to itself) the
diagonal entries are redundant. Moreover, because connectivity is a symmetric relation (i.e. if
node A is connected to node B then node B is connected to node A) the connectivity can be tested
once for each pair of nodes and hence only half of the possible route inspections are required.
Consequently, what is needed is only to verify that node $n$ is connected to all nodes $m$ ($m>n$)
where $n=1,2,\ldots,(N-1)$ with $N$ being the total number of nodes. Hence, the square array
reduces to a lower or upper triangular array.

In fact even this is a stronger condition than is needed because a necessary and sufficient
condition for total connectivity is that a single randomly-selected node (let us call it X) is
connected to all other nodes. The reason is that connectivity is a transitive relation (i.e. if A
is connected to B and A is connected to C then B is connected to C) and hence under this reduced
connectivity condition any two nodes in the network are connected to each other at least through
node X. The triangular array then reduces to a one-dimensional array whose entries represent
connectivity status of node X with respect to all other nodes in the network (e.g. one row
representing node X versus $N-1$ columns representing all other nodes).

However, for an unstructured network even this reduced route inspection is not an easy task in its
simple and direct realization, and hence other methods which are implicitly based on this method
are easier in implementation and more efficient in execution, as we will see for example in the
following node mapping method.

\subsection{Node Mapping}

The idea of node mapping, whose essence is the reduced route inspection as outlined in the last
section, is that instead of testing the connectivity between two pre-selected nodes, which is very
difficult and demanding since there are many possibilities for the routes and junctions between
these two particular nodes, the search is focused on finding all the connections between random
nodes with a start from a single randomly-selected node. By compiling these random connections into
a list of connected nodes, a connected partition, which could incorporate the entire network, can
be constructed. The starting point in this method is to build a node-to-neighbors mapping (i.e. a
nodes list that maps each node to its immediate neighbor nodes) and provide it as an input. This
can take the form of a structured indexed array whose first entry is a node index while its second
entry is a vector of the indices of all neighbor nodes. An implicit node index can be used for the
sake of efficiency and reduced memory storage and hence the first entry is redundant. It is very
easy and efficient to assemble such a mapping from the network segments list where the index of
each node of a segment is added to the list of connected nodes in the entry of the other node.

Node mapping starts from a randomly-selected node by adding this node and all its neighbors to a
connected partition list with the removal of the node itself from the list of network nodes. The
process then goes on recursively by adding the neighbor nodes of each one of the previously found
neighbors with the removal of these neighbor nodes from the network nodes list as soon as their
immediate neighbors are added to the neighbor list and hence to the current connected partition.
The process ends either with the complete consumption of the nodes list if the network is totally
connected, or with the failure of finding any new connected node in one iteration which marks the
identification of a complete connected partition. In the latter case, this operation can be
repeated, to find the other partitions, until the network nodes list is exhausted.

\begin{figure}[!h]
\centering{}
\includegraphics
[scale=0.8] {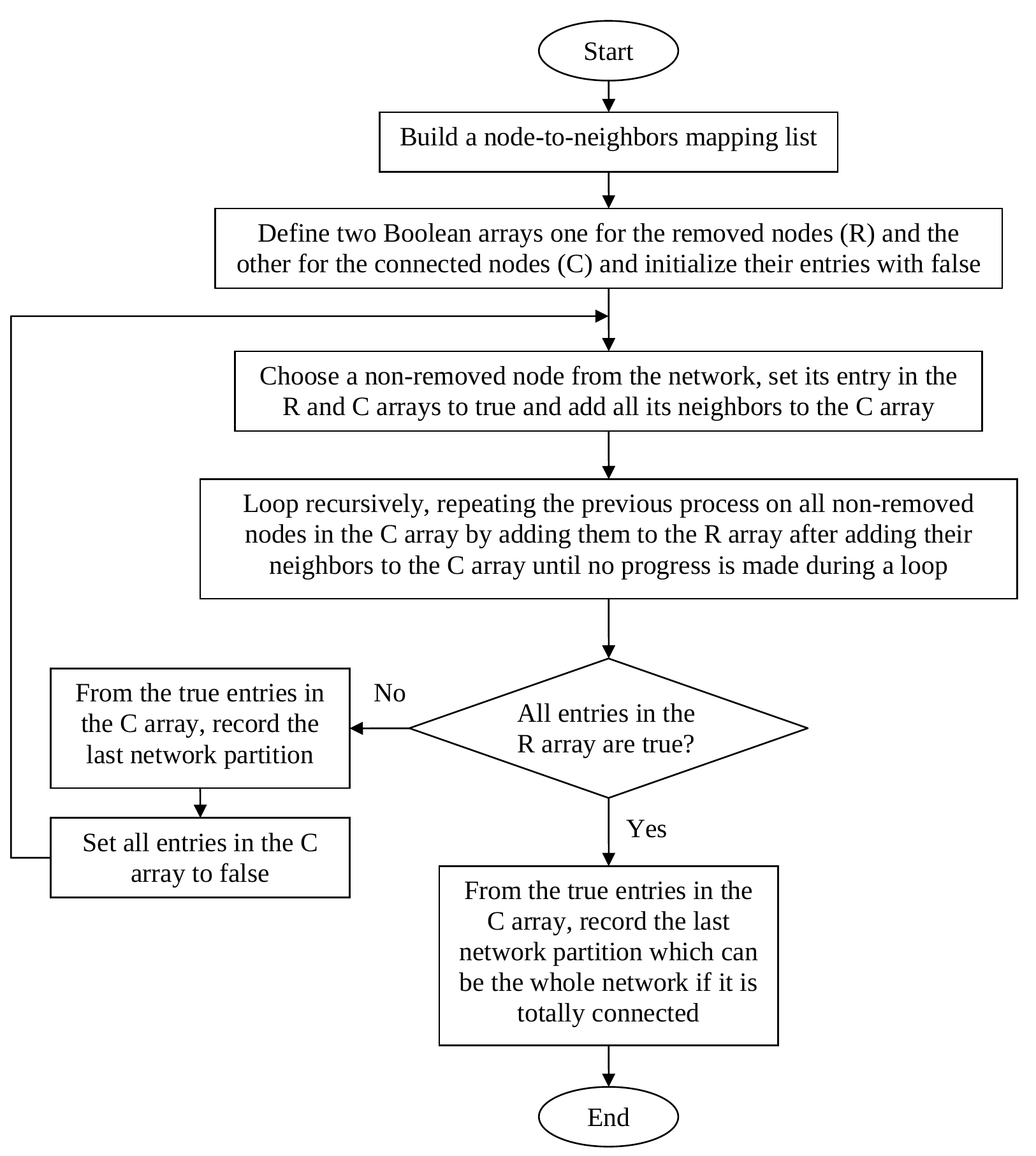} \caption{Generic flow chart of an algorithm based on the node mapping method.}
\label{FlowNM}
\end{figure}

The use of two 1D Boolean arrays, one to mark the removal of the nodes from the network list and
the other to mark the affiliation of the nodes to the current partition, can make this process
highly efficient in terms of memory, especially if implicit node indices (represented by the
ordinal indices of the array entries) are used for accessing the entries of the Boolean arrays,
with an added advantage of ease of implementation. The direct access of the Boolean entries of the
nodes by the prompt use of their indices will speed up the whole operation and make it highly
efficient in the CPU time as well. A flow chart of a possible algorithmic implementation based on
the node mapping method with the use of two 1D Boolean arrays is presented in Figure \ref{FlowNM}.
Other computational techniques, such as the actual removal of the processed nodes from the nodes
list directly, can also be used for implementing this method. However, some of these techniques may
incur a high computational cost in terms of CPU time and memory consumptions and could complicate
the implementation.

The use of direct access through the use of node indices may be complicated by the existence of
missing node indices as the network indexing is assumed to be random. This may also be caused by
the absence of some indices due, for example, to the removal of singular nodes. However, this can
be easily managed by re-indexing the nodes orderly to remove the missing indices, at least
temporarily while performing the connectivity test. The nodes can be re-indexed back to their
original indices after completing the connectivity test. This difficulty can also be overcome
through the use of large storage with a vacant entries for the missing indices. These vacant
entries can be marked (e.g. by filling them with a certain invalid value) so that they are excluded
from the involvement in the connectivity test. The latter may incur an unnecessary large storage
especially for very large networks if the missing indices create large indices gaps.

\subsection{Segment Mapping}

In this highly efficient method, the search for connectivity starts from seeding a list of
connected nodes by the two nodes of a randomly selected segment. By going through the remaining
segments and adding the node of any segment whose other node is found on the connected nodes list,
a connected partition, which possibly comprises the whole network, will gradually build up. All
segments whose nodes are added to the list are removed from the segments list either directly or by
the use of a labeling mechanism such as a 1D Boolean array to mark the status of the segments as
being removed or not. The inspection of the segments list is repeated until the segments list is
empty (in which case the network is totally connected) or the inspection of all the remaining
segments in the list in one of the iteration cycles returns no new nodes to be added to the
connected nodes list (in which case the network is dismembered and partially connected). In the
latter case, this inspection process can be repeated iteratively to identify all the partitions of
the network until the exhaustion of the entire segments list.

\begin{figure}[!h]
\centering{}
\includegraphics
[scale=0.8] {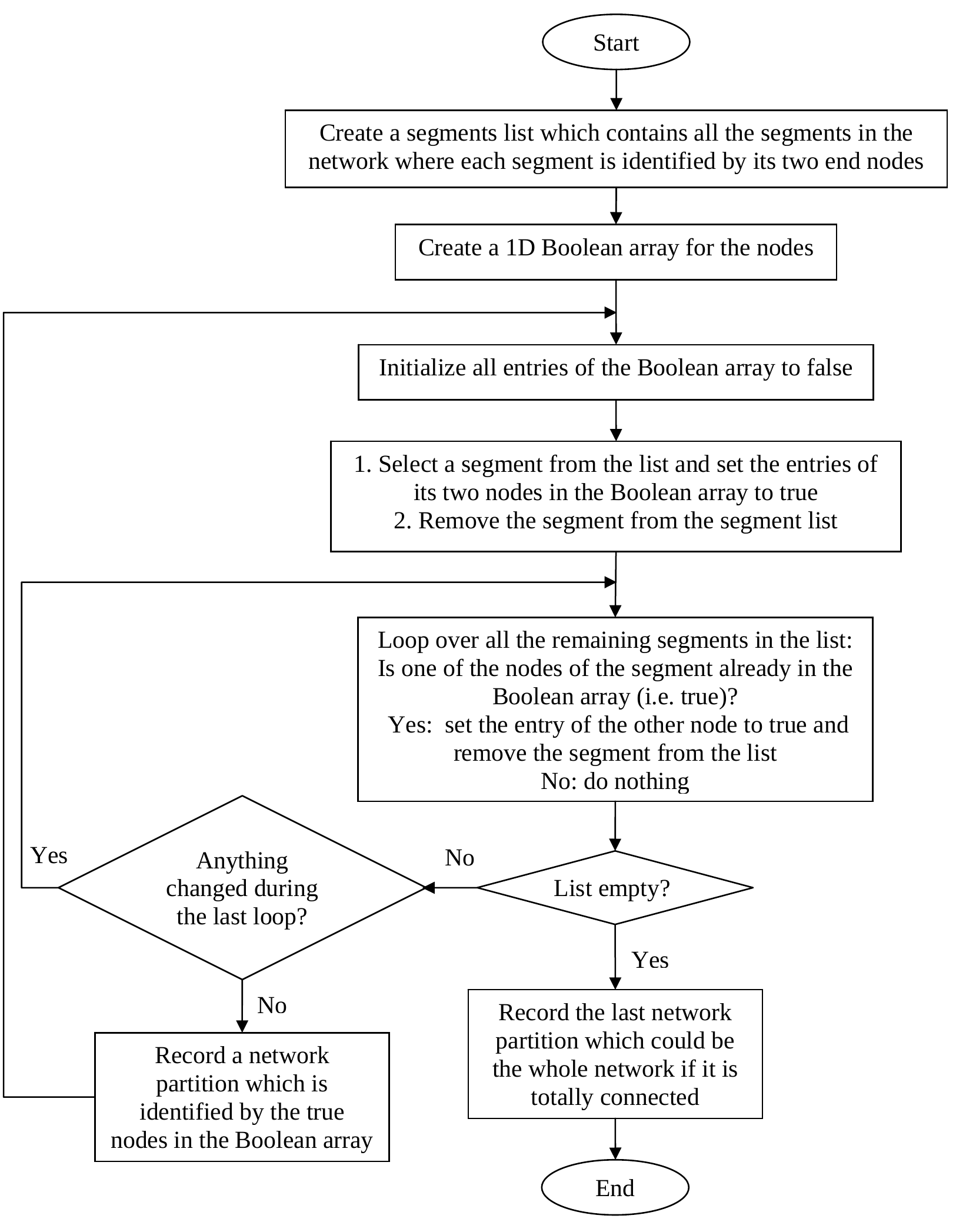} \caption{Generic flow chart of an algorithm based on the segment mapping
method.} \label{FlowSM}
\end{figure}

The input data required for this method is a list of the network segments where each segment is
identified by the indices of its two end nodes. The efficiency of this method can be enhanced by
employing a direct access technique to the nodes status (as being included in the current partition
or not) through possible use of an indexed array without need for searching a nodes list. The use
of a 1D Boolean array, whose implicit cell index can be used to indicate the node index, can
therefore facilitate the marking of the nodes if they are connected to the current partition or
not. All the entries of this array may be initially set to false at the start of each partition
search, and the entries of the connected nodes can be set to true as the search goes on. By the end
of each partition search, the nodes that belong to that partition (which possibly include the whole
network) can be identified from the implicit indices of the true cells. A similar technique can
also be used to identify all the segments that belong to each partition. A flow chart of a possible
computational algorithmic implementation based on the segment mapping method with the use of a 1D
Boolean array for labeling the nodes and recording their status with respect to the current
partition is presented in Figure \ref{FlowSM}.

Other techniques for recording the nodes connected to the partition, such as adding these nodes and
their segments to a network partition array, can also be used instead of the use of a 1D Boolean
array. These techniques may be used to provide more detailed information about the partition, such
as the segments of the partition as well as its nodes, with minimum post processing requirements
although they should require more memory consumption and processing time.

\section{Comparison}

In this section, we present a general comparison between the node mapping and segment mapping
methods and their associated algorithms with their advantages and shortcomings. However, we would
like to insist that most of the conclusions derived from these comparisons are dependent on the
particular implementation of these algorithms.

\begin{figure}[!h]
\centering{}
\includegraphics
[scale=0.6] {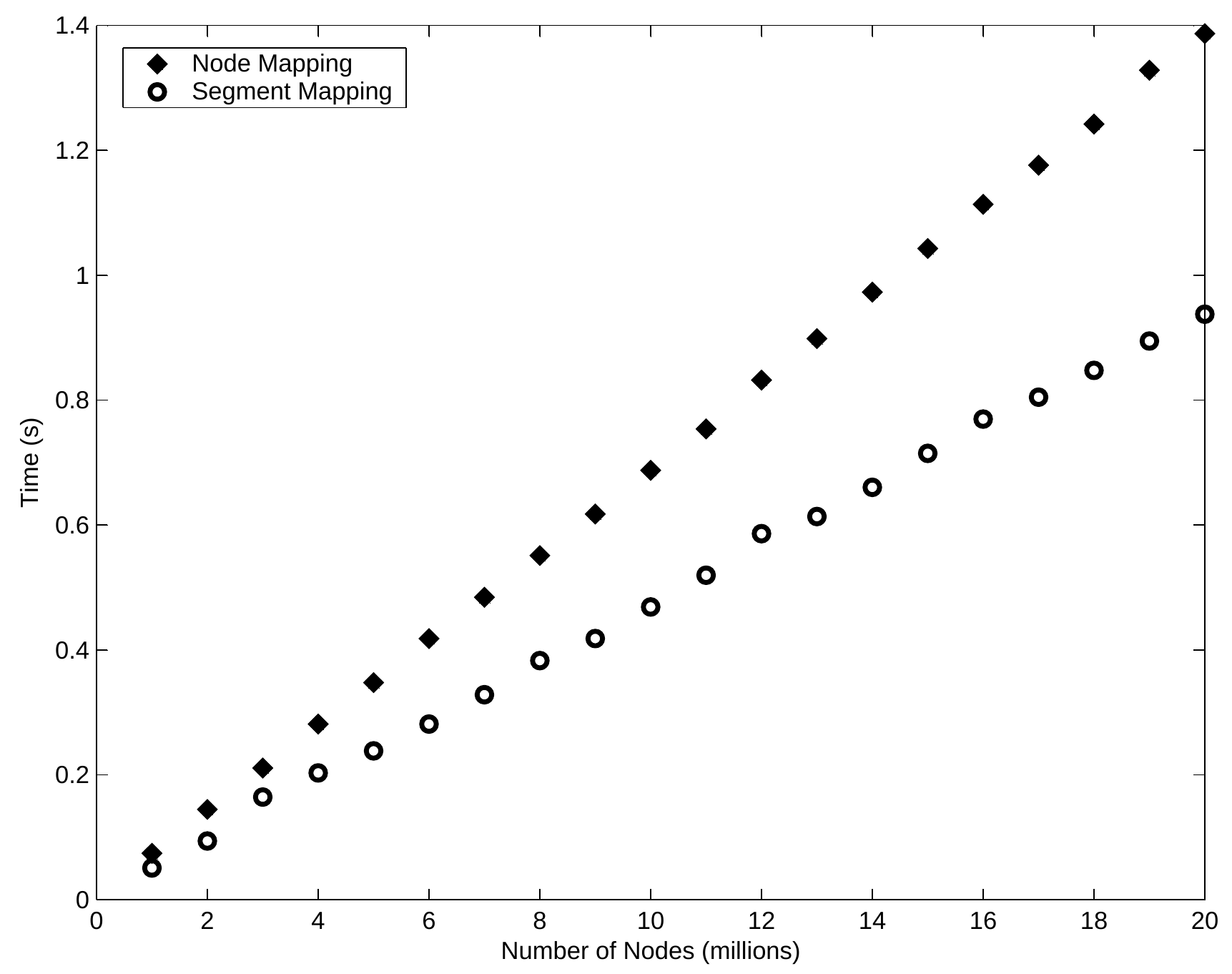} \caption{Execution time in seconds versus the number of network nodes in
millions for the node mapping and segment mapping algorithms. All networks used in these tests have
a single partition with an average connectivity index of approximately 5.} \label{Comp1}
\end{figure}

\begin{figure}[!h]
\centering{}
\includegraphics
[scale=0.6] {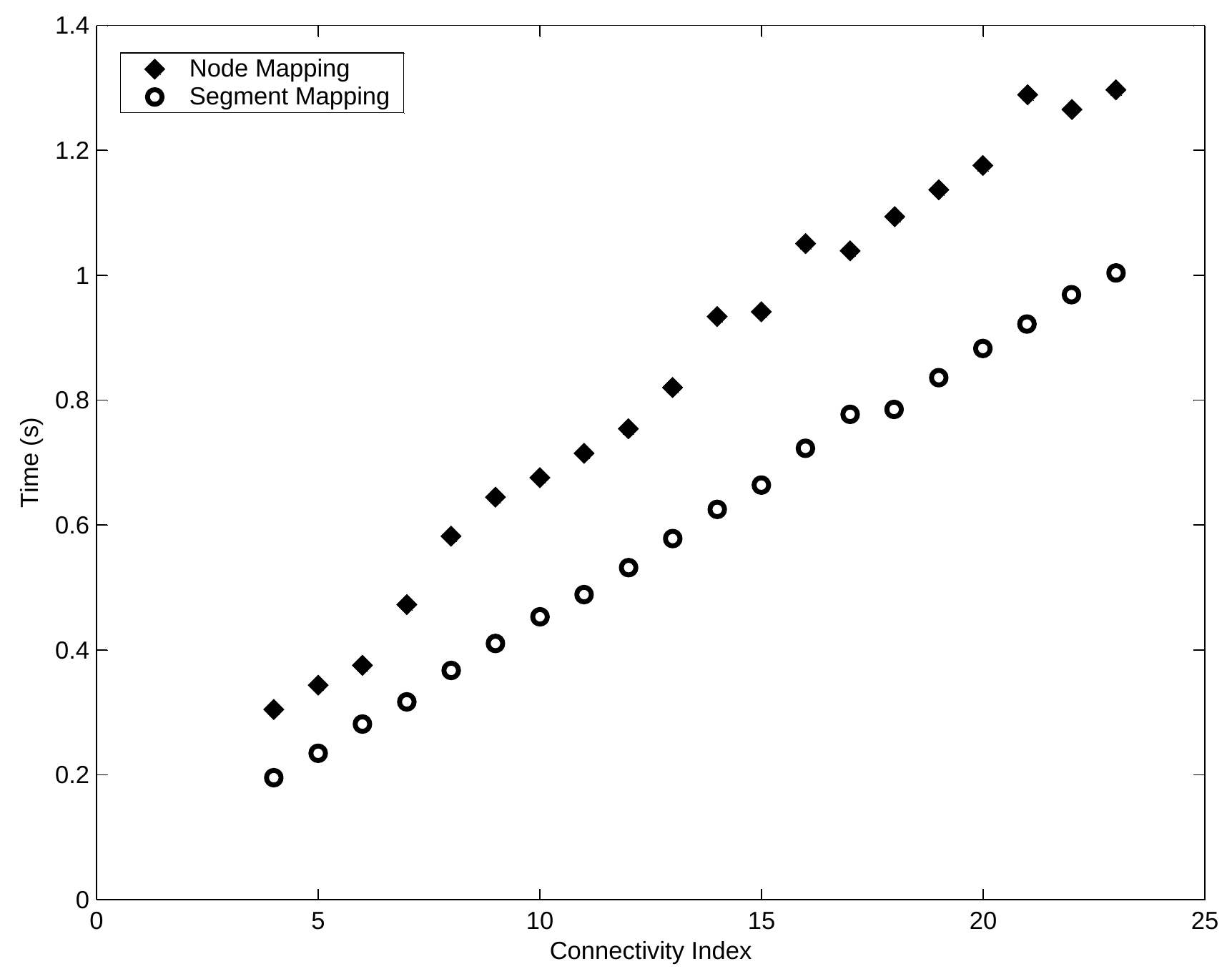} \caption{Execution time in seconds versus the average connectivity index of the
network for the node mapping and segment mapping algorithms. All networks used in these tests have
a single partition with five million nodes.} \label{Comp2}
\end{figure}

\begin{figure}[!h]
\centering{}
\includegraphics
[scale=0.6] {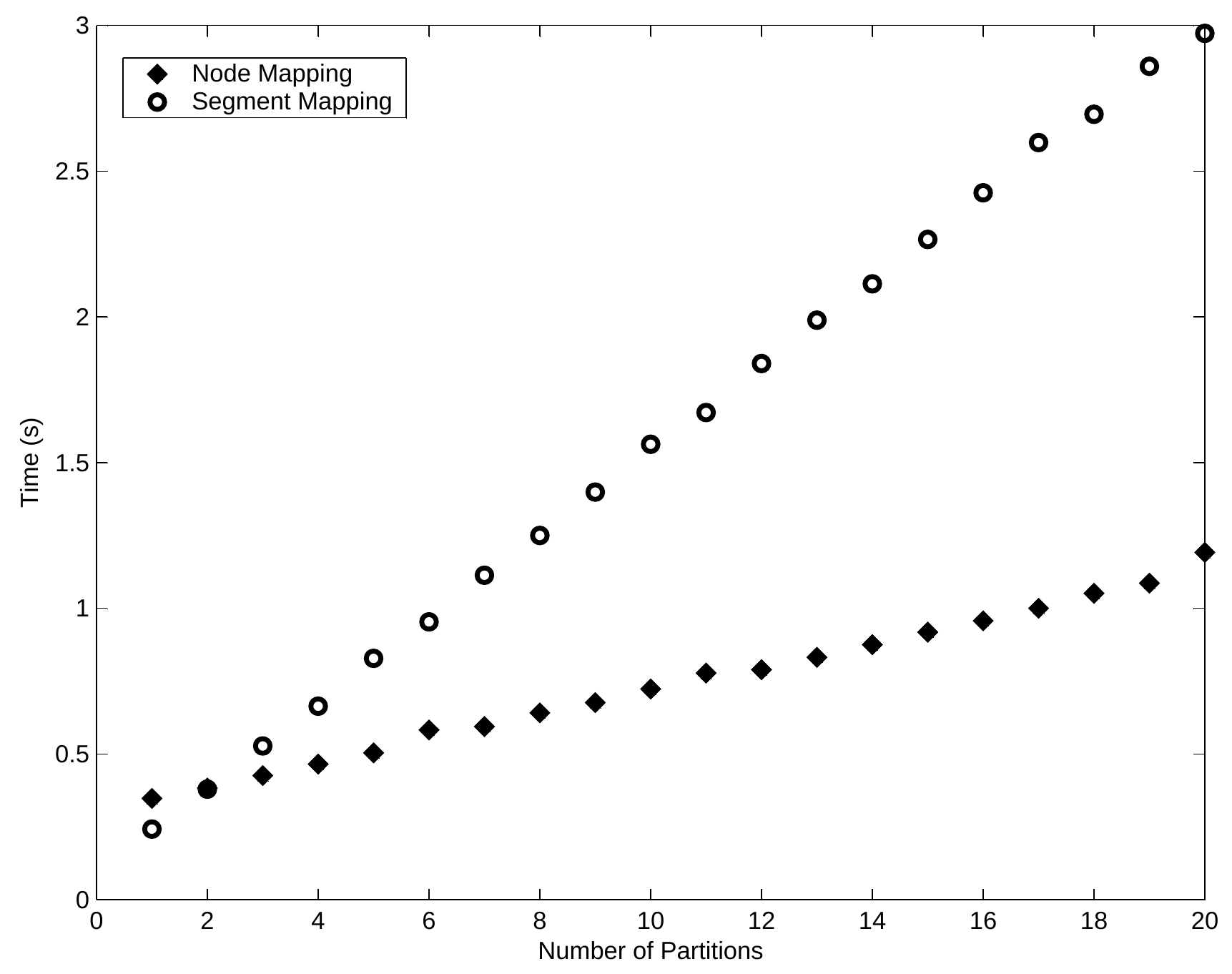} \caption{Execution time in seconds versus the number of network partitions for
the node mapping and segment mapping algorithms. All networks used in these tests have about five
million nodes with an average connectivity index of approximately 5.} \label{Comp3}
\end{figure}

The node mapping and segment mapping algorithms, as outlined in the flow charts of Figures
\ref{FlowNM} and \ref{FlowSM}, were implemented in a C++ computer program. Many tests have been
carried out; some of these are presented in Figures \ref{Comp1}, \ref{Comp2} and \ref{Comp3}. In
these figures, the average execution time from several runs has been taken to smooth out
fluctuations. In all these tests, a Visual Studio 6.0 compiler on a normal laptop computer with a
1.99 GHz processor and 1.87 GB of RAM memory running under Windows XP operating system was used.
All the networks utilized in these tests are computer generated using a stochastic computational
procedure based on establishing random connections between randomly selected nodes with certain
constraints on the number of nodes, connectivity index and number of partitions.

As seen in Figures \ref{Comp1}, \ref{Comp2} and \ref{Comp3}, although the segment mapping has a
better performance with respect to the network size (as quantified by the number of nodes and
number of segments which is related to the average connectivity index) when the network is totally
connected, the performance of the node mapping becomes superior for partitioned networks.

The memory requirement of the node mapping, as outlined in the flow chart of Figure \ref{FlowNM},
is $2N$ Boolean storage plus $Nc_{av}$ integer storage where $N$ is the total number of nodes and
$c_{av}$ is the average connectivity index of the network. For the segment mapping, the memory
requirement, as outlined in the flow chart of Figure \ref{FlowSM}, is $N$ Boolean storage plus $2M$
integer storage where $M$ is the number of segments. Because $2M=Nc_{av}$, the segment mapping
requires less memory according to the implementation of Figure \ref{FlowSM}. However, if the
removal of segments from the segments list is achieved not directly but through the use of a
segments Boolean array to mark the removed segments, then an extra $M$ Boolean storage is required
for the segment mapping, and hence its memory requirement will exceed the requirement of the node
mapping when $c_{av}>2$, which is typically the more common case. Anyway, the memory costs for both
algorithms are affordable with modest computational resources even for very large networks.

An advantage of the segment mapping is that more detailed information about the network partitions
(that is information about both nodes and segments of partitions) can be obtained directly with no
need for extra computational effort. In node mapping, what is found directly is the nodes of each
partition, and hence extra effort is required to obtain information about the segments in these
partitions.

\clearpage
\section{Conclusions}

The importance of networks in many scientific and engineering applications cannot be overstated.
Therefore, testing and validation of network connectivity is crucial to obtain valid computational
results. In this article, two highly optimized methods, node mapping and segment mapping, for
verifying network connectivity with their associated computer algorithms have been presented. Both
methods have their merits. However, the main advantage of both is that they can be easily
implemented and used for testing the connectivity and acquiring the network partitions.

As implemented, segment mapping is superior in terms of speed of execution for totally connected
networks, while node mapping is superior for partitioned networks. The superiority of these
algorithms with regard to the memory storage requirement depends on the network characteristics,
such as average connectivity index, as well as the particular algorithmic implementation. Other
factors that affect the performance of these methods include the type of network and its topology,
the type of iteration used to loop over the nodes and segments, connectivity index distribution,
and the order of storing the nodes and segments in their arrays which depends on the indexing.

According to the author's implementation of the node mapping and segment mapping algorithms, both
memory and time of execution scale linearly with the size of the network as quantified by the
number of nodes and number of segments which is correlated to the average connectivity index of the
network (refer to Figures \ref{Comp1} and \ref{Comp2}). The time of execution also scales linearly
with the number of partitions (refer to Figure \ref{Comp3}).

\newpage
\phantomsection \addcontentsline{toc}{section}{References} %
\bibliographystyle{unsrt}

\begin{thebibliography}{10}

\bibitem{SorbieCJ1989}
K.S. Sorbie; P.J. Clifford;~E.R.W. Jones.
\newblock {The Rheology of Pseudoplastic Fluids in Porous Media Using Network
  Modeling}.
\newblock {\em Journal of Colloid and Interface Science}, 130(2):508--534,
  1989.

\bibitem{OrenBA1997}
P.E. {\O}ren; S. Bakke;~O.J. Amtzen.
\newblock {Extending Predictive Capabilities to Network Models}.
\newblock {\em SPE Annual Technical Conference and Exhibition, San Antonio,
  Texas, SPE 38880}, 1997.

\bibitem{TsakiroglouTKPS2003}
C.D. Tsakiroglou; M. Theodoropoulou; V. Karoutsos; D. Papanicolaou;~V. Sygouni.
\newblock {Experimental study of the immiscible displacement of shear-thinning
  fluids in pore networks}.
\newblock {\em Journal of Colloid and Interface Science}, 267(1):217--232,
  2003.

\bibitem{PerrinTSC2006}
C.L. Perrin; P.M.J. Tardy; S. Sorbie;~J.C. Crawshaw.
\newblock {Experimental and modeling study of Newtonian and non-Newtonian fluid
  flow in pore network micromodels}.
\newblock {\em Journal of Colloid and Interface Science}, 295(2):542--550,
  2006.

\bibitem{Sochithesis2007}
T.~Sochi.
\newblock {\em {Pore-Scale Modeling of Non-Newtonian Flow in Porous Media}}.
\newblock PhD thesis, Imperial College London, 2007.

\bibitem{SochiB2008}
T.~Sochi;~M.J. Blunt.
\newblock {Pore-scale network modeling of Ellis and Herschel-Bulkley fluids}.
\newblock {\em Journal of Petroleum Science and Engineering}, 60(2):105--124,
  2008.

\bibitem{SherwinFPP2003}
S.J. Sherwin; V. Franke; J. Peir\'{o};~K. Parker.
\newblock {One-dimensional modelling of a vascular network in space-time
  variables}.
\newblock {\em Journal of Engineering Mathematics}, 47(3-4):217--250, 2003.

\bibitem{RuanCZC2003}
W.~Ruan; M.E. Clark; M. Zhao;~A. Curcio.
\newblock {A Hyperbolic System of Equations of Blood Flow in an Arterial
  Network}.
\newblock {\em SIAM Journal on Applied Mathematics}, 64(2):637--667, 2003.

\bibitem{FormaggiaLTV2006}
L.~Formaggia; D. Lamponi; M. Tuveri;~A. Veneziani.
\newblock {Numerical modeling of 1D arterial networks coupled with a lumped
  parameters description of the heart}.
\newblock {\em Computer Methods in Biomechanics and Biomedical Engineering},
  9(5):273--288, 2006.

\bibitem{AlastrueyMPDPS2007}
J.~Alastruey; S.M. Moore; K.H. Parker; T. David; J. Peir\'{o}~S.J. Sherwin.
\newblock {Reduced modelling of blood flow in the cerebral circulation:
  Coupling 1-D, 0-D and cerebral auto-regulation models}.
\newblock {\em International Journal for Numerical Methods in Fluids},
  56(8):1061--1067, 2008.

\bibitem{RouxH1987}
S.~Roux;~A. Hansen.
\newblock {A new algorithm to extract the backbone in a random resistor
  network}.
\newblock {\em Journal of Physics A}, 20:L1281--L1285, 1987.

\bibitem{Schilders2009}
W.H.A. Schilders.
\newblock {Predicting the topology of dynamic neural networks for the
  simulation of electronic circuits}.
\newblock {\em Neurocomputing}, 73(1-3):127--132, 2009.

\bibitem{ZhanMT2009}
S.~Zhan; J.F. Miller;~A.M. Tyrrell.
\newblock {An evolutionary system using development and artificial Genetic
  Regulatory Networks for electronic circuit design}.
\newblock {\em Biosystems}, 98(3):176--192, 2009.

\bibitem{RatnayakeJ1999}
N.~Ratnayake;~I.N. Jayatilake.
\newblock {Study of transport of contaminants in a pipe network using the model
  EPANET}.
\newblock {\em Water Science and Technology}, 40(2):115--120, 1999.

\bibitem{HammaDF2008}
V.~Hamma; P. Collon-Drouaillet;~R. Fabriol.
\newblock {Two modelling approaches to water-quality simulation in a flooded
  iron-ore mine (Saizerais, Lorraine, France): A semi-distributed chemical
  reactor model and a physically based distributed reactive transport pipe
  network model}.
\newblock {\em Journal of Contaminant Hydrology}, 96(1-4):97--112, 2008.

\bibitem{DobersekG2009}
D.~Dobersek;~D. Goricanec.
\newblock {Optimisation of tree path pipe network with nonlinear optimisation
  method}.
\newblock {\em Applied Thermal Engineering}, 29(8-9):1584--1591, 2009.

\bibitem{BentleyL1980}
R.W. Bentley;~T.A. Lambe.
\newblock {Assignment of traffic to a network of signalized city streets}.
\newblock {\em Transportation Research Part A: General}, 14(1):57--65, 1980.

\bibitem{MessmerP1994}
A.~Messmer;~M. Papageorgiou.
\newblock {Automatic control methods applied to freeway network traffic}.
\newblock {\em Automatica}, 30(4):691--702, 1994.

\bibitem{Ledoux1997}
C.~Ledoux.
\newblock {An urban traffic flow model integrating neural networks}.
\newblock {\em Transportation Research Part C: Emerging Technologies},
  5(5):287--300, 1997.

\bibitem{SarvothamRB2005}
S.~Sarvotham; R. Riedi;~R. Baraniuk.
\newblock {Network and user driven alpha-beta on-off source model for network
  traffic}.
\newblock {\em Computer Networks}, 48(3):335--350, 2005.

\bibitem{JiangXCHX2011}
D.~Jiang; Z. Xu; Z. Chen; Y. Han;~H. Xu.
\newblock {Joint time-frequency sparse estimation of large-scale network
  traffic}.
\newblock {\em Computer Networks}, 55(15):3533--3547, 2011.

\bibitem{SohY1998}
B.C. Soh;~S. Young.
\newblock {Network system and world wide web security}.
\newblock {\em Computer Communications}, 20(16):1431--1436, 1998.

\bibitem{BarabasiAJ2000}
A.-L. Barab\'{a}si; R. Albert;~H. Jeong.
\newblock {Scale-free characteristics of random networks: the topology of the
  world-wide web}.
\newblock {\em Physica A: Statistical Mechanics and its Applications},
  281(1-4):69--77, 2000.

\bibitem{Sgroi2008}
D.~Sgroi.
\newblock {Social network theory, broadband and the future of the World Wide
  Web}.
\newblock {\em Telecommunications Policy}, 32(1):62--84, 2008.

\bibitem{CrucittiLM2004}
P.~Crucitti; V. Latora;~M. Marchiori.
\newblock {A topological analysis of the Italian electric power grid}.
\newblock {\em Physica A: Statistical Mechanics and its Applications},
  338(1-2):92--97, 2004.

\bibitem{ChassinP2005}
D.P. Chassin;~C. Posse.
\newblock {Evaluating North American electric grid reliability using the
  Barab\'{a}si-Albert network model}.
\newblock {\em Physica A: Statistical Mechanics and its Applications},
  355(2-4):667--677, 2005.

\bibitem{LinY2011}
Y.-K. Lin; C.-T. Yeh.
\newblock {Maximal network reliability with optimal transmission line
  assignment for stochastic electric power networks via genetic algorithms}.
\newblock {\em Applied Soft Computing}, 11(2):2714--2724, 2011.

\bibitem{HuismanBD2002}
T.~Huisman; R.J. Boucherie;~N.M. van Dijk.
\newblock {A solvable queueing network model for railway networks and its
  validation and applications for the Netherlands}.
\newblock {\em European Journal of Operational Research}, 142(1):30--51, 2002.

\bibitem{MeesterM2007}
L.E. Meester;~S. Muns.
\newblock {Stochastic delay propagation in railway networks and phase-type
  distributions}.
\newblock {\em Transportation Research Part B: Methodological}, 41(2):218--230,
  2007.

\bibitem{SochiEmis2010}
T.~Sochi.
\newblock {Emissivity: A program for atomic transition calculations}.
\newblock {\em Communications in Computational Physics}, 7(5):1118--1130, 2010.

\end{thebibliography}

\end{document}